\documentclass[runningheads]{llncs}

\usepackage{graphicx}
\usepackage{amsmath}
\usepackage{booktabs}
\usepackage{graphicx}
\usepackage{comment}
\usepackage{mdwlist}
\usepackage{paralist}
\usepackage{tabularx}
\usepackage{soul}
\usepackage{adjustbox}
\usepackage[colorinlistoftodos]{todonotes}
\usepackage{hyperref}
\usepackage{subfig}
\usepackage{float}
\usepackage{algorithm2e}
\usepackage{xcolor} 
\usepackage{enumitem}
\usepackage{todonotes}
\usepackage{multirow}
\usepackage{rotating}
\usepackage{colortbl}
\usepackage{hyperref}
\usepackage{pifont}
\usepackage{xspace}
\usepackage{color,soul}
\usepackage{common}
\usepackage{booktabs}
\usepackage{siunitx}

\usepackage[show]{chato-notes}

\definecolor{magicmint}{rgb}{0.67, 0.94, 0.82}
\definecolor{melon}{rgb}{0.99, 0.74, 0.71}
\definecolor{ashgrey}{rgb}{0.7, 0.75, 0.71}

\usepackage{pgfplots,capt-of}
\pgfplotsset{compat=1.9}
\usepgfplotslibrary{external}
\newcommand{\para}[1]{\paragraph{\textnormal{\textbf{#1}.}}}
\newcommand{\parawodot}[1]{\paragraph{\textnormal{\textbf{#1}}}}

\newcommand{\argminH}{\mathop{\mathrm{argmin}}\limits}

\newcommand{\qppbert}{QPP-SRF}
\newcommand{\mcf}{TD2F}
\newcommand{\afr}{LR-SRF}
\newcommand{\darf}{Deep-SRF}

\newcommand{\darfbert}{Deep-SRF-BERT}
\newcommand{\rrf}{R2F2}

\newcommand{\QE}{\phi_M(Q)}
\renewcommand{\vec}[1]{\mathbf{#1}}
\newcommand{\defas}{\overset{\mathrm{def}}{=\joinrel=}}

\newcommand{\uls}{\begin{itemize}[leftmargin=*]}
\newcommand{\ule}{\end{itemize}}
\newcommand{\ols}{\begin{enumerate}[leftmargin=*]}
\newcommand{\ole}{\end{enumerate}}

\begin{document}

\title{A Deep Learning Approach for Selective Relevance Feedback}

\author{
Suchana Datta
\inst{1}
\orcidID{0000-0001-9220-6652}
\and
Debasis Ganguly
\inst{2}
\orcidID{0000-0003-0050-7138} \and
Sean MacAvaney
\inst{2}
\orcidID{0000-0002-8914-2659} \and
Derek Greene
\inst{1}
\orcidID{0000-0001-8065-5418}
}
\authorrunning{S. Datta et al.}
%
\institute{University College Dublin, Ireland \and
University of Glasgow, United Kingdom \\
\email{suchana.datta@ucdconnect.ie},
\email{debasis.ganguly@glasgow.ac.uk},
\email{sean.macavaney@glasgow.ac.uk},
\email{derek.greene@ucd.ie}}

\maketitle

\pagestyle{empty}

\begin{abstract}
Pseudo-relevance feedback (PRF) can enhance average retrieval effectiveness over a sufficiently large number of queries. However, PRF often introduces a drift into the original information need, thus hurting the retrieval effectiveness of several queries. While a selective application of PRF can potentially alleviate this issue, previous approaches have largely relied on unsupervised or feature-based learning to determine whether a query should be expanded. In contrast, we revisit the problem of selective PRF from a deep learning perspective, presenting a model that is entirely data-driven and trained in an end-to-end manner. The proposed model leverages a transformer-based bi-encoder architecture. Additionally, to further improve retrieval effectiveness with this selective PRF approach, we make use of the model's confidence estimates to combine the information from the original and expanded queries. In our experiments, we apply this selective feedback on a number of different combinations of ranking and feedback models, and show that our proposed approach consistently improves retrieval effectiveness for both sparse and dense ranking models, with the feedback models being either sparse, dense or generative.


  
\end{abstract}

\section{Introduction}
\label{sec:intro}
\begin{figure}[!t]
\centering

\includegraphics[width=.95\columnwidth]
{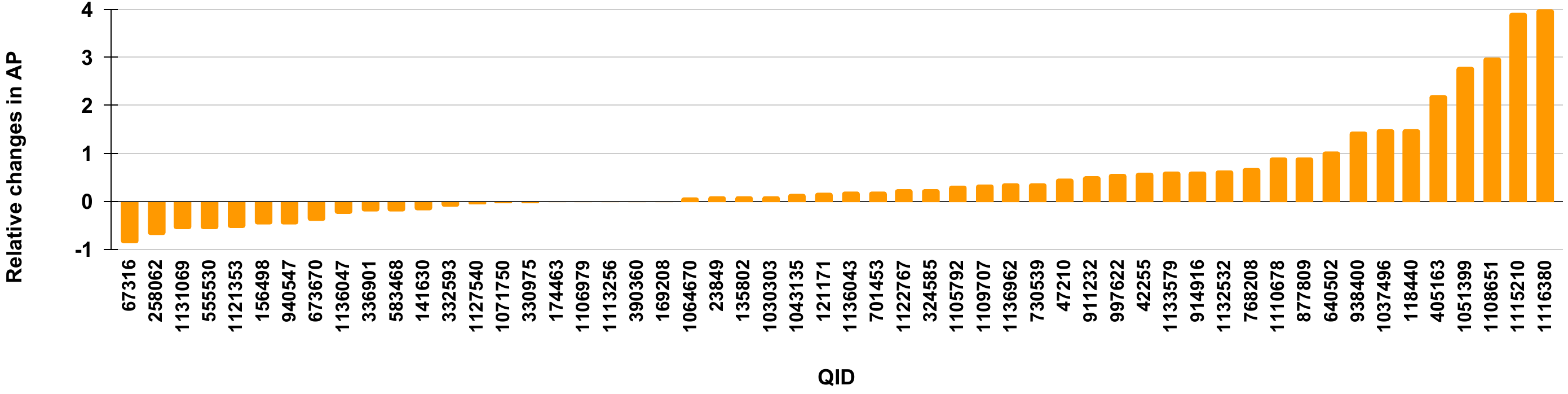}
\caption{
\small
Relative changes in AP, i.e., (AP(post-fdbk) - AP(pre-fdbk))/AP(pre-fdbk), for TREC DL'20 queries. We observe that many queries are negatively impacted by PRF (bars below the x-axis).}
\label{fig:barchart}
\end{figure}

The keywords that a user enters as query to a search engine are often insufficient to express the user's information need, resulting in a \textit{lexical gap} between the text in the query and the relevant documents~\cite{belkin1982ask}. Standard pseudo-relevance feedback (PRF) methods, such as the relevance model \cite{Lavrenko_RLM2001:RBL:383952.383972} and its variants \cite{DBLP:conf/coling/GangulyLJ12,NMF_ProbabilisticMatrixFactorization,DoiKDERLM_CIKM16,generative-rlm}, can overcome this problem and ultimately yield improvements in retrieval effectiveness.
Generally speaking, PRF methods
are designed to enrich a user's initial query with distinctive terms from the top-ranked documents \cite{rocchio,automaticQE,lca}.
Despite the demonstrated success of PRF in improving retrieval effectiveness, a number of studies have identified certain limitations of this strategy~\cite{questioningQueryExpansion,zhai-cikm09,croft-selective-exp,josiane_prf_tois}. For the most part, these limitations share a common theme: there is no consistent PRF setting that works well across a wide range of queries; to put in simple words,
\emph{one size does not fit all}. Figure \ref{fig:barchart} illustrates such a situation, where nearly $38.9\%$ of queries from TREC DL'20 topic set are penalized as a result of PRF.
It has been shown that not all documents contribute equally well to PRF, 
as certain documents may impair retrieval effectiveness when used to expand a query \cite{cluster-resample,kNNbasedPRF_Bashir_CIKM09}. This can even be true when relevant documents are used to enrich a query's representation \cite{PoisonPills}. It has also been observed that
some queries are amenable to more aggressive query expansion, while others work better with more conservative settings \cite{ellen-qe}. Moreover, not all terms might contribute equally well in terms of enriching the representation of a query \cite{goodExpansionTerm_SIGIR2008,find-good-fdbk-doc}, which suggests that a selective approach to PRF can potentially improve the overall IR effectiveness.

\begin{figure}[!t]
\centering
\includegraphics[width=.85\columnwidth]{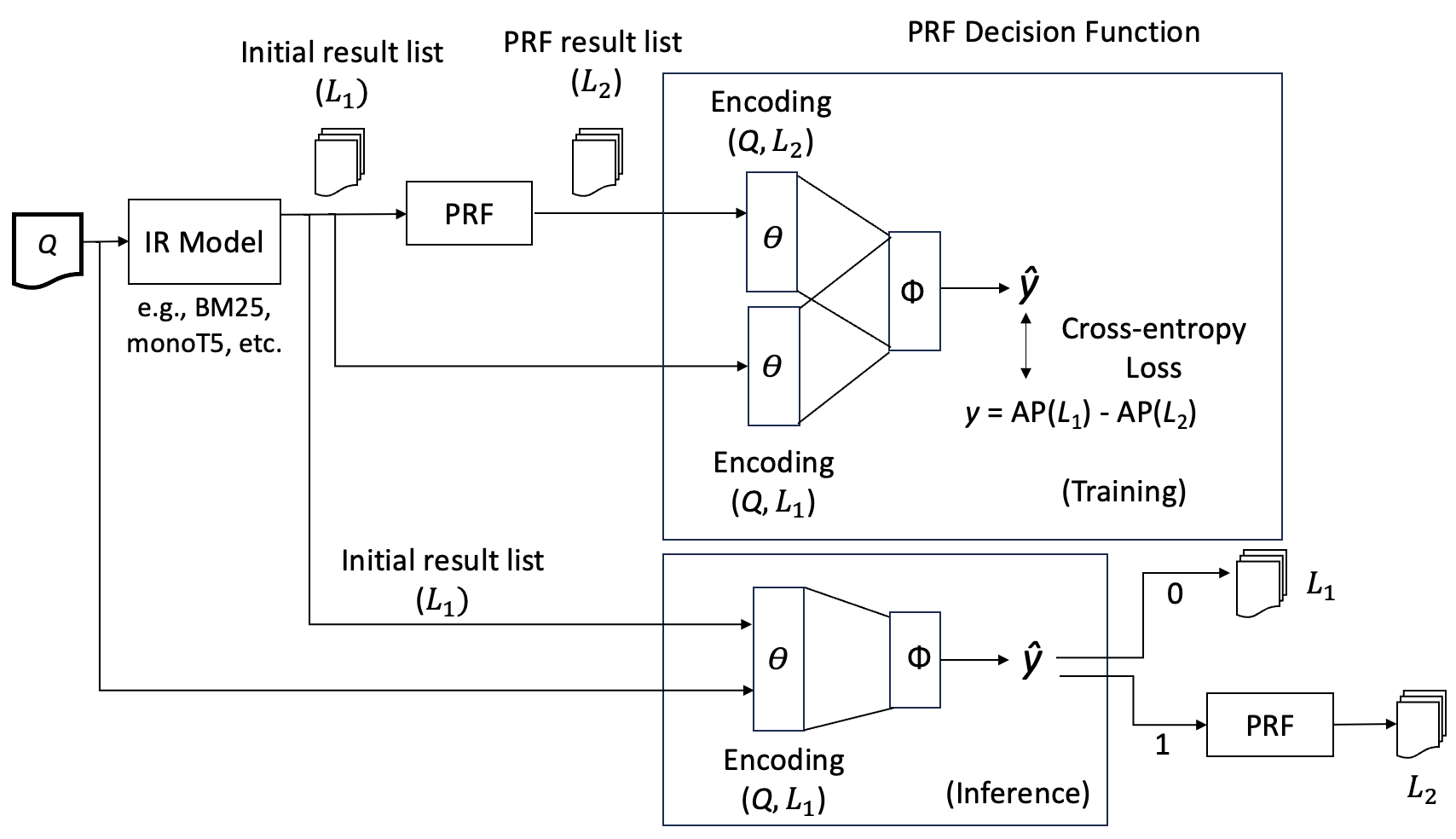}
\caption{
\small
A schematic diagram of selective feedback. The main contribution of this paper is a supervised data-driven approach towards realising the decision function.}
\label{fig:intro_schematic}
\end{figure}

Rather than following the previous approaches on
adapting the number of feedback terms \cite{ellen-qe} or attempting to choose a robust subset of documents for PRF \cite{cluster-resample,kNNbasedPRF_Bashir_CIKM09}, we rather focus on solving the more fundamental decision question of
``\emph{whether or not to apply PRF for a given query}'' \cite{croft-selective-exp,zhai-cikm09} with the help of a supervised data-driven approach.
We hypothesise that selectively applying feedback to only those queries that are amenable to PRF can improve the overall retrieval effectiveness by avoiding query drift in cases where feedback would not be beneficial. 
Our idea is depicted in Figure \ref{fig:intro_schematic}.




The main novelty of our proposed selective pseudo relevance feedback (SRF) approach is that in contrast to existing work on selective PRF, we propose a data-driven supervised neural model for predicting which queries are conducive to PRF. More specifically, during the training phase we make use of the relevance assessments to learn a decision function that, given the query and the top-retrieved set of documents both with and without feedback, predicts whether it is useful to apply PRF. During the inference phase, we make use of only a part of the shared parameter network which, given a query and its top-retrieved document set, predicts whether PRF is to be applied (schematically illustrated in Figure \ref{fig:intro_schematic}).
This way of inference reduces computational costs for queries where PRF should eventually be ignored.


A key advantage of our SRF approach is that it can be applied to the output ranked list obtained by \textbf{any retrieval model}, ranging from sparse models (e.g., BM25, LM-Dir etc.) to dense ones (e.g., MonoBERT \cite{monobert}).
Moreover, in the SRF workflow it also is possible to use \textbf{any PRF model} to enrich a query's representation, ranging from sparse models (e.g., RLM \cite{Lavrenko_RLM2001:RBL:383952.383972}) to dense ones (e.g., ColBERT-PRF \cite{colbert-prf}) to even generative ones (e.g., GRF \cite{generative-rlm}).

\section{Related Work} \label{sec:review}



The evolution of relevance feedback in IR spans from traditional query expansion models \cite{ellen-qe,goodExpansionTerm_SIGIR2008} to cluster-based feedback document selection \cite{cluster-resample,find-good-fdbk-doc}. While prior research has considered both unsupervised selective feedback \cite{croft-selective-exp} and feature-driven methods \cite{zhai-cikm09}, we introduce a data-driven neural strategy for selective relevance feedback. Several existing methods, both supervised and unsupervised, hinge on \emph{decision-based relevance feedback}. One unsupervised approach uses Query Performance Prediction (QPP) scores \cite{kurland_tois12,wig_croft_SIGIR07,clarity-croft,uef_kurland_sigir10,iadh-ipm07}, which we include as a baseline. The higher the QPP score, the greater the chance of identifying relevant documents at the top rank positions with the initial query. However, high variances in retrieval status values, as seen in neural re-rankers like MonoBERT \cite{monobert}, can make QPP scores deceptive. To avoid such heuristics, our method focuses solely on query terms and the documents retrieved by that query in order to learn the selection function.

\para{PRF on and for Dense Retrieval}

Recently, the community has seen a significant interest in feedback for dense retrieval to boost performance. Precursors to dense feedback models made use of word embeddings for PRF, e.g., KDERLM \cite{DoiKDERLM_CIKM16} which proposed a generalised RLM with word embeddings, and PRF-NMF \cite{ZamaniDSC16}, which leveraged matrix factorisation to bridge the semantic gap between terms from a query and its top-retrieved documents.  

The study by \cite{jamie-cikm21} explored relevance feedback principles within dense retrieval models. Li et al. \cite{prf_signal_guido} analyzed feedback signal quality, comparing traditional models like Rocchio \cite{rocchio} with dense retrievers like ANCE-based retrievers \cite{ance-iclr}, finding the latter more resilient. Representation models, such as ColBERT \cite{colbert_sigir20}, can allow us to append additional embedding layers to the query representation, as demonstrated by \cite{xiao-ictir21}. This method employed contextualized PRF to cluster and rank feedback document embeddings in order to select suitable expansion embeddings, thus improving document ranking.  In other work, \cite{counterfactual_prf_guido} leveraged implicit feedback from historical clicks for relevance feedback in dense retrieval.  The authors introduced counterfactual-based learning-to-rank, showing that historic clicks can be highly informative in terms of relevance feedback. Lastly, \cite{guido_interpolate_sigir22} proposed the idea of fusing feedback signals from both sparse and dense retrievers in the context of PRF.  

More recently, PRF on dense IR models has garnered significant interest \cite{nprf,ceqe,bertqe,colbert-prf}. The concept of `dense for PRF' was first emphasized in \cite{reinforce_prf}, which proposed a reinforcement-based learning algorithm designed to explore and exploit various retrieval metrics, aiming to learn an optimized PRF function. With the recent success of LLMs, \cite{generative-rlm} proposed a generative feedback method (GRF) that makes use of LLM generated long-form texts instead of first pass retrieved results to build a probabilistic feedback model. In contrast, our work aims to develop a generic PRF strategy that does not apply feedback blindly, but rather learns a selection function in a supervised manner to analyze the suitability of relevance feedback for each query irrespective of sparse or generative PRF.

\para{Selective PRF}
Prior work on selective PRF has considered either fully unsupervised approaches \cite{croft-selective-exp} or feature-based supervised approaches \cite{zhai-cikm09} for selective relevance feedback (SRF).
The former makes use of query performance prediction (QPP) based measures
to predict if a query should be expanded, where the decision depends on whether the QPP score exceeds a given threshold. On the other hand, existing supervised approaches first represent each query as a bag of characteristic features derived from its top-retrieved set of documents. 
A classifier is then trained on these features to predict whether or not a query should be expanded \cite{zhai-cikm09}.

\section{Model Description} \label{sec:modeloverview}
\subsection{A Generic Decision Framework for PRF} \label{ss:genericmodel}

Given a set of queries $\mathcal{Q}=\{Q_1,\ldots,Q_n\}$, a standard relevance feedback model $M$ uses the information from the top-retrieved documents of each query to enrich its representation, i.e., $M: Q \mapsto \phi_M(Q)$. Consequently, each query $Q \in \mathcal{Q}$ is transformed to an enriched representation $\phi_M(Q)$, which is then used either for re-ranking the initial list, or to execute a second-step retrieval.

Unlike the standard PRF setting, a decision-based selective PRF framework first applies a \emph{decision function}, $\theta: Q \mapsto \{0, 1\}$, which outputs a Boolean to indicate if the retrieval results for $Q$ is likely to be improved after application of PRF. As per our proposal, the overall PRF process on the set of queries $\mathcal{Q}$ does not simply replace each query $Q$ with its enriched form $\phi_M(Q)$. Rather, it makes use of the function $\theta(Q)$ for each query $Q$ to decide whether to output the initial ranked list or to make use of the enriched query representation $\phi_M(Q)$, as obtained by a PRF model $M$ (leading to either re-ranking the initial list or re-retrieving a new list via a second stage retrieval). 
The top-$k$ ranked list of documents, $L_k(Q) = \{D^Q_1,\ldots,D^Q_k\}$, retrieved for a query $Q$, in addition to being a function of the query $Q$ itself, is thus also a function of i) the feedback model $M$, ii) the enriched query representation $\phi_M(Q)$, and iii) the decision function $\theta$, i.e.,
\begin{equation}
L_k(Q) =
\begin{cases}        
\sigma(Q), & \text{if $\theta(Q)=0$}\\
\sigma(\phi_M(Q)), & \text{if $\theta(Q)=1$,}
\end{cases}
\label{eq:selPRFgeneric}
\end{equation}
where $\sigma(Q)$ denotes a retrieval model, e.g., BM25, that outputs an ordered set of $k$ documents sorted by the similarity scores.

Previous approaches have explored the use of both unsupervised and supervised approaches for addressing this decision problem. We now briefly explain both strategies in our own context.

\para{Unsupervised decision function}
An unsupervised approach, such as \cite{croft-selective-exp}, applies a threshold parameter on a QPP estimator function, $\theta_{\mathrm{QPP}}: Q \mapsto [0,1]$.
More concretely, if the predicted QPP score is lower than the threshold parameter, it is likely to indicate that the retrieval performance for the query has scopes for further improvement and subsequently PRF is applied for this query.
Formally speaking, the decision function of an unsupervised approach takes the form
\begin{equation}
\theta(Q) \defas \mathbb{I}(\theta_{\mathrm{QPP}} < \tau) \label{eq:unsupdecision},
\end{equation}
where $\tau \in [0, 1]$ is the threshold parameter.

\para{Supervised decision function} 
In the supervised approach, the decision function also depends on the enriched query representation and its top-retrieved documents.
More precisely, a supervised PRF decision is
a parameterized function of features of i) the query $Q$, ii) its top-retrieved documents $L_k(Q)$, iii) the enriched query $\phi_M(Q)$, and iv) its top-retrieved set $L_k(\phi_M(Q))$ \cite{zhai-cikm09}.
%
The training process itself makes use of a set of queries $\mathcal{Q}_{\text{train}}$ for which ground-truth indicator labels are computed by evaluating the relative effectiveness obtained with the original query vs. the enriched query with the help of available relevance assessments. Formally,
\begin{equation}
y(Q) = \mathbb{I}(\mathrm{AP}(\phi_M(Q)) > \mathrm{AP}(Q)),
\label{eq:refindicators}
\end{equation}
where $\mathrm{AP}(Q)$ denotes the average precision of a query $Q \in \mathcal{Q}_{\text{train}}$.
The indicator values of $y(Q)$ are used to learn the parameters of a classifier function to yield a supervised version of the decision function $\theta$:
\begin{equation}
\theta(Q) \defas \zeta\cdot \vec{z}_{Q,\phi_M(Q)},\,\, \text{where}\,\,
\theta(Q) \approx \argminH_{\zeta}\!\!\!\! \sum_{Q' \in \mathcal{Q'}_{\text{train}}}\!\!\!\!(y(Q') - \zeta\cdot \vec{z}_{Q',\phi_M(Q')})^2.
\label{eq:classifier}
\end{equation}
In Equation \ref{eq:classifier}, $\zeta$ represents a set of learnable parameters, with $\vec{z}_{Q',\phi_M(Q')}$ denoting a set of features extracted from both the original query $Q'$ and the enriched query $\phi_M(Q')$ along with the features from their top-retrieved set of documents $L_k(Q')$ and $L_k(\phi_M(Q'))$.
The variable $y(Q')$, as per the definition of $y(Q)$, denotes the ground-truth indicating if PRF should be applied for $Q'$.
The optimal parameter vector $\zeta$, as learned from a training set of queries $\mathcal{Q}_{\text{train}}$ (Equation \ref{eq:classifier}) is then used to predict the decision for any new query $Q$. The features we use are described later in the paper in Section \ref{ss:baseline}. 
%



\subsection{Transformer-based Encoding for PRF Decision}

We now describe a data-driven approach for learning the decision function with deep neural networks. 
%
Instead of making use of a specific set of extracted features as used in the QPP model in \cite{datta_deepqpp22}, the learning objective of a transformer-based PRF model makes use of the terms present in the documents and the queries. As with Equation \ref{eq:classifier}, we make use of both the content of the original query $Q$ and its enriched form $\phi_M(Q)$, along with their top-retrieved sets.
More formally,
\begin{equation}
\theta(Q) \defas
\zeta\cdot(
\mathcal{E}(Q,D^Q_1,\ldots,D^Q_k) \oplus
\mathcal{E}(\phi_M(Q),D^{\QE}_1,\ldots,D^{\QE}_k)
),
\end{equation}
where $\theta(Q)$ is learned by computing
\begin{equation}
\argminH_{\zeta}\!\!\!\!\! \sum_{Q' \in \mathcal{Q'}_{\text{train}}}\!\!\!\!\!\!(y(Q') -
\zeta\cdot(\mathcal{E}(Q',L_k(Q')) \oplus \mathcal{E}(\phi_M(Q'),L_k(\phi_M(Q'))))^2.
\label{eq:classifierdatadriven}
\end{equation}
In Equation \ref{eq:classifierdatadriven}, $\mathcal{E}$ is a parameterized function for encoding the interaction between a query $Q$ and its top-retrieved documents, $L_k(Q)$. This encoding function, generally speaking, maps a query (a sequence of query terms) and a sequence of documents (which are themselves sequences of their constituent terms) into a fixed length vector, i.e., $\mathcal{E}: Q, L_k \mapsto \mathbb{R}^p$ ($p$ an integer, e.g., for BERT embeddings $p=768$). Here $\oplus$ indicates an interaction operation (e.g., a merge layer in a neural network) between the query-document encodings corresponding to the original query and the enriched one. 

\begin{figure}[t]
\centering
\includegraphics[width=.85\columnwidth]{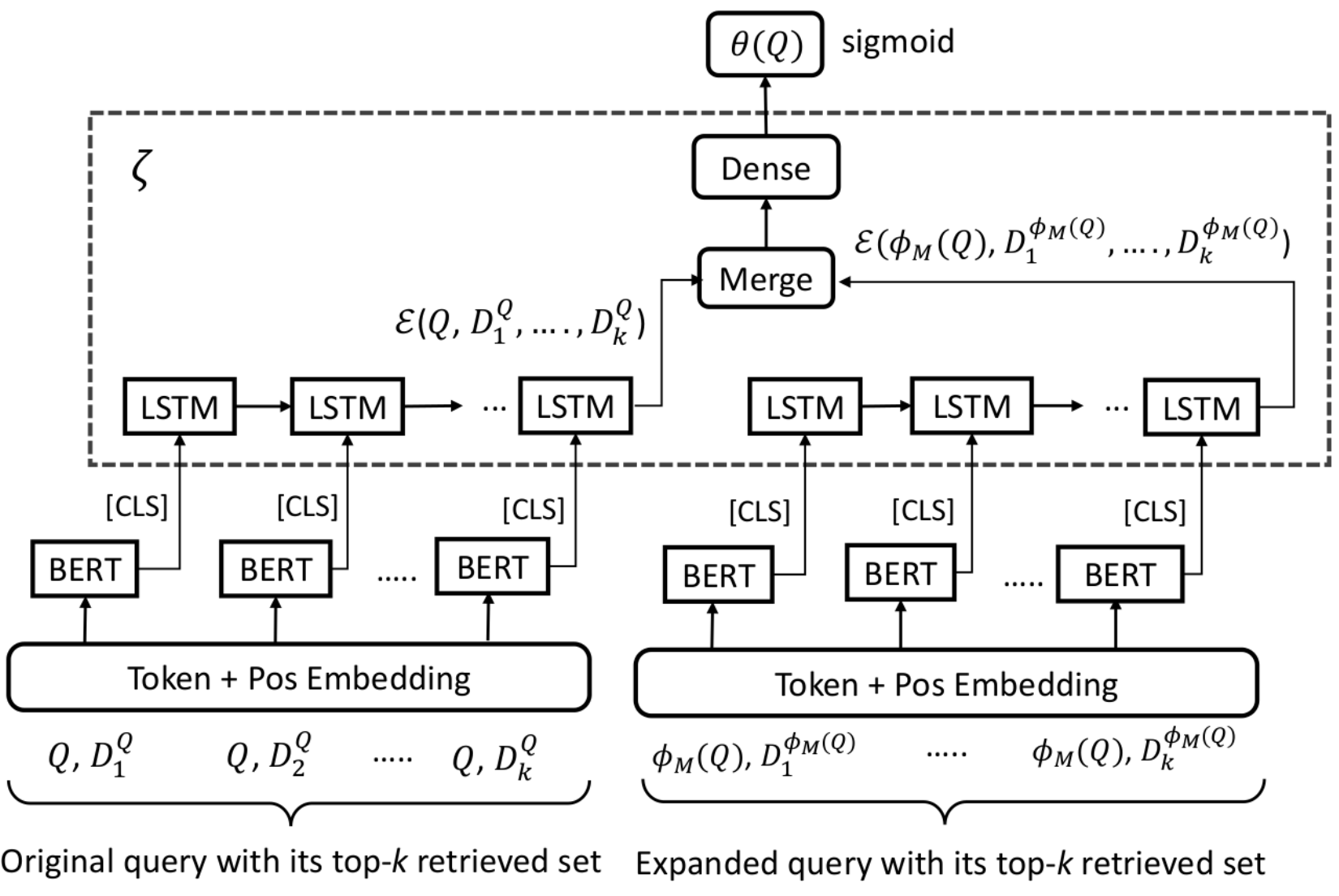}
\caption{
\small
Training of a transformer based query-document architecture with shared parameters for selective PRF. During inference, only the left part of the network is used to output whether to apply PRF or not for a given query. 
}
\label{fig:bert}
\end{figure}
The transformer-based encoding uses the BERT architecture which takes as input the contextual embeddings of the terms for each pair comprising a query $Q$ and its top-retrieved document $D^Q_i \in L_k(Q)$.
The 768 dimensional `[CLS]' representations of each \emph{query-document} pair is then encoded with LSTMs as a realisation of the encoded representation of a query and its top-retrieved set, i.e., to define $\mathcal{E}(Q, L_k(Q))$ as per the notation of Equation \ref{eq:classifierdatadriven}.
We further obtain a BERT-based encoding of the expanded query $\phi_M(Q)$ and its top-retrieved set and then merge the two representations before passing them through a feed-forward network. More formally,
\begin{equation}
\mathcal{E}(\bar{Q}, L_k(\bar{Q}))=
\text{LSTM}(
\text{BERT}(\bar{Q},D^{\bar{Q}}_1)_{[\text{CLS}]},\ldots,
\text{BERT}(\bar{Q},D^{\bar{Q}}_k)_{[\text{CLS}]}
).
\label{eq:bertenc}
\end{equation}
The variable $\bar{Q} \in \{Q, \phi_M(Q)\}$, i.e., in one branch of the network it corresponds to the original query, whereas in the other it corresponds to the expanded one. Figure \ref{fig:bert} shows the transformer-specific implementation of the encoding function. The set of learnable parameters $\zeta$ in this case comprises of the LSTM and the fully connected (dense) layer parameters, as shown in Figure \ref{fig:bert}.  
We name this particular model \textbf{Deep-SRF-BERT} (Deep Selective Relevance Feedback with the use of BERT transformers).

\para{Model Inference} During inference, only the part corresponding to the original query and its top-retrieved set of documents is used to predict the output variable (a sigmoid) which if higher than 0.5 indicates that PRF should be applied.

\subsection{Model Confidence-based PRF Calibration}

Prior work has applied confidences of prediction models to improve retrieval effectiveness \cite{DBLP:conf/sigir/CohenMLRE21}.
In our work, we use the uncertainties in the prediction of the decision function to further improve search results.
Rather than only reporting either results with or without relevance feedback, we make use of the confidence of the decision function $\theta(Q)$ to combine the results from the two lists -- one without feedback and the other with feedback. Specifically, if the supervised model outlined in Section \ref{ss:genericmodel} is decisive in its choice between $L_k(Q)$ (original query retrieved list) and $L_k(\QE)$ (the list retrieved with the expanded query), then one of the rankings is expected to dominate over the other. 
However, when the model $\theta(Q)$ itself is not confident about the prediction, we can potentially achieve better results if we `meet somewhere in the middle'.

Formally, we propose a rank-fusion based method, where the fusion weights are obtained from the predictions of the PRF decision model $\theta(Q)$. The predicted value $\theta(Q)$ being a sigmoid represents the probability of classifying the decision into one of the two outcomes - the closer $\theta(Q)$ is to $0$, the higher is the model's confidence in not applying feedback, and similarly the closer $\theta(Q)$ is to $1$, the higher is the model's confidence in applying PRF. The predicted value of $\theta(Q) \in [0,1]$ can thus be used as weights to fuse the two different ranked lists, i.e., the fusion score $\sigma_F(Q,D)$ of a document $D$ for a query $Q$ is given by
\begin{equation}
\sigma_F(Q, D)=\frac{1-\theta(Q)}{\mathrm{Rank}(D,L_k(Q))} + \frac{\theta(Q)}{\mathrm{Rank}(D,L_k(\QE))} \label{eq:rankfusion},
\end{equation}
where the notation $\mathrm{Rank}(D,L)$ denotes the rank of a document $D$ in a list $L$.

If $D \notin L$, then the rank is set to a large value $\aleph (>k)$, which in our experiments was set to $1000$ (higher than all possible values of $k$ we experimented with). For values of $\theta(Q)$ close to $0.5$ (the most uncertainty in prediction), the fusion-based approach leads to a more uniform contribution from both the lists. In contrast, a value of $\theta(Q)$ close to $0$ ensures that the majority of the score contribution comes from the original query (because $1-\theta(Q) >> \theta(Q)$), and a similar argument applies for $\theta(Q)\rightarrow 1$, in which case the major contribution comes from the second term in the right-hand side of Equation \ref{eq:rankfusion}.



\section{Evaluation} \label{sec:setup}


\subsection{Research Questions}


Since a primary contribution of this paper is the idea of applying a fully data-driven approach, the first research question that we investigate is whether a shift from the existing feature-based approach for selective PRF to a data-driven one does indeed result in improved retrieval effectiveness. Therefore, we formulate our first research question as follows.

\noindent\textbf{- RQ1:} Does SRF lead to overall improvements in IR effectiveness over non-selective and other baseline approaches?






Our second research question aims to investigate whether the model prediction uncertainty-based fusion of the two ranked lists -- one retrieved with the original query and the other with the expanded one, can potentially improve the retrieval effectiveness further.

\noindent\textbf{- RQ2:} Can we use the confidence estimates from our selective PRF model for a \emph{soft selection} of information from both pre-feedback and post-feedback sources to further improve IR effectiveness?



In our third research question, the aim is to investigate how effectively the selection strategy in PRF \textit{transfers} across different feedback approaches, i.e., training the SRF approach only once on a PRF model (e.g., RLM), and then apply the decision model on other PRF models (e.g., ColBERT-PRF).

\noindent\textbf{- RQ3:} Does selection effectively transfers the learning across different PRF approaches?

\subsection{Methods Investigated} \label{ss:baseline}


We consider a range of unsupervised and supervised methods, described below.
Some baselines refer to existing methods, while others are extensions of alternative approaches to allow us to provide a fair comparison, such as by using a more recent QPP method instead of the originally-proposed clarity score \cite{croft-selective-exp}. 




\parawodot{PRF} is a standard non-selective relevance feedback model, namely the relevance model (RLM) \cite{Lavrenko_RLM2001:RBL:383952.383972}. We use the RM3 version of the relevance model as reported in \cite{NasreenJaleel_RM3}, which is a linear combination of the weights of the original query model and new expansion terms.
In fact, we use RLM as one of the base PRF model $M$ 
which means that the standard RLM degenerates to a specific case of the generic selective PRF framework of Equation \ref{eq:selPRFgeneric} with $\theta(Q)=1\,\forall Q \in \mathcal{Q}$, i.e., when for each query we use its enriched form $\QE$.



\parawodot{\rrf}
refers to an adaptation of the Reciprocal Rank-based Fusion (RRF) \cite{fusion-clarke}, a simple yet effective approach for combining the document rankings from multiple IR systems. For our task, instead of combining ranked lists from two different retrieval models, we merge the ranked lists of the original and the expanded queries, i.e., $L_k(Q)$ and $L_k(\QE)$ as per our notations. We name the adapted method Reciprocal Rank Fusion-based Feedback (R2F2).
%
%
%
Formally, the score for document $D$ after fusion is given by
\begin{equation}
\sigma_F(Q, D)=\frac{1-\alpha}{\mathrm{Rank}(D,L_k(Q))} + \frac{\alpha}{\mathrm{Rank}(D,L_k(\QE))} \label{eq:r2f2},
\end{equation}
where, similar to Equation \ref{eq:rankfusion} $\mathrm{Rank}(D, L)$ denotes the rank of a document in a list $L$ (this being a large number $\aleph$ if $D \notin L$), and $\alpha\in [0,1]$ is a linear combination hyper-parameter that we adjust with grid search on each training fold. A lower value of $\alpha$ puts more emphasis on the initial retrieval list, whereas a higher value ensures that the feedback rank of a document contributes more. 
Equation \ref{eq:r2f2} is
a special case of Equation \ref{eq:rankfusion} with a constant value of $\theta(Q)=\alpha$ for each query $Q$. 

\parawodot{QPP-SRF}
is an adaptation of the method proposed in \cite{croft-selective-exp}, where the QPP score of a query is used as estimate to decide if PRF should be applied for that query (see $\theta(Q)$ in Section \ref{ss:genericmodel}). The idea here is that a high QPP score is already indicative of an effective retrieval performance, in which case, the method avoids any further risk of potentially degrading the retrieval quality with query expansion.
We refer to this method as QPP-based selective relevance feedback (QPP-SRF). The method requires a base QPP estimator to provide $\theta_{\mathrm{QPP}}$ scores.
%

To choose the QPP estimator, we conducted a set of initial experiments using several standard unsupervised QPP approaches.
The recently introduced supervised QPP method qppBERT-PL \cite{datta-qppbertpl} demonstrated the best downstream retrieval effectiveness.
Therefore, we report results of QPP-SRF combined with qppBERT-PL, where training is conducted using the settings as reported in \cite{datta-qppbertpl}.
%
%
A key parameter for QPP-SRF is the threshold value $\tau$ ($\tau \in [0, 1]$) which controls the decision around whether PRF is applied or not. In our experiments we tune $\tau$ on the train folds. To ensure that the threshold can be applied for any QPP estimate, we normalize the QPP estimates in the range $[0, 1]$.

\parawodot{\mcf}
is an unsupervised selective feedback approach that is conceptually similar to QPP-SRF \cite{croft-selective-exp}. Rather than using a QPP method, it computes the difference of the term weight distributions across the sets of documents retrieved with the original and the expanded queries, i.e., the sets $L_k(Q)$ and $L_k(\QE)$ as per our notations introduced in Section \ref{ss:genericmodel}. Formally,
\begin{equation}
\theta(Q) = \frac{1}{|V|}\sum_{t\in V}\log P(t|L_k(Q))-\log P(t|L_k(\QE)),
\end{equation}
where the set $V$ denotes the vocabulary of the two lists, i.e., $V = V_{L_k}\cup V_{L_k(\QE)}$.
As per \cite{croft-selective-exp}, we set the feedback decision threshold $\tau$ to a value such that
over $95\%$ of the queries satisfy the criterion that $\theta(Q) \leq \tau$. We name this method as Term Distribution Divergence based Feedback, or \mcf~for short.




\begin{table}[t]
\centering
\caption{
\small
Summary of the data used in our experiments. The columns `$\bar{|Q|}$' and `$\bar{\#Rel}$' denote average number of query terms and average number of relevant documents.}

\begin{adjustbox}{width=0.85\columnwidth}
\small
\begin{tabular}{@{}l@{~~~~}r@{~~~~}r@{~~~~}r@{~~~~}r@{~~~~}r@{}}
\toprule
Collection & \#Docs & Topic Set & \#Topics & $\bar{|Q|}$ & $\bar{\#Rel}$ \\

\midrule



\multirow{3}{*}{MS MARCO Passage} 
& \multirow{3}{*}{8,841,823} 
& MS MARCO Train & 502,939 & 5.97 & 1.06 \\
& & TREC DL'19 & 43 & 5.40 & 58.16 \\
& & TREC DL'20 & 54 & 6.04 & 30.85 \\

\bottomrule
\end{tabular}

\label{tab:dataset}
\end{adjustbox}
\end{table}

\parawodot{\afr}
is the only existing supervised method  that uses the query features, 
along with their top-retrieved documents, to predict the PRF decision \cite{zhai-cikm09}. The ground-truth labels for learning the decision function 
is obtained for a training set of queries with existing relevance assessments, i.e. $y(Q) = \mathbb{I}(\mathrm{AP}(\phi_M(Q)) > \mathrm{AP}(Q))$. The method then uses Equation \ref{eq:classifier} to train a feature-based logistic regression classifier.
In particular, the experiments reported in \cite{zhai-cikm09} used the following features for training the logistic regression model: i) the clarity \cite{clarity-croft} of top-retrieved documents, ii) the absolute divergence between the query model $Q$ and the relevance model \cite{Lavrenko_RLM2001:RBL:383952.383972}, iii) the Jensen-Shannon divergence \cite{shanon-div} between the language model of the feedback documents, and iv) the clarity of the query language model.
We name this method as Regression-based Selective relevance Feedback (\afr).


\parawodot{Proposed methods}
In addition to conducting experiments with our proposed model \darfbert~(Figure \ref{fig:bert}), we also incorporate confidence-based calibration (as per objective \textbf{RQ2}) with rank fusion (Equation \ref{eq:rankfusion} and \ref{eq:r2f2}), which we denote by adding the suffix \rrf\footnote{Implementation available at: \url{https://github.com/suchanadatta/AdaptiveRLM.git}}.

\subsection{Experimental Setup}

\para{Dataset and train-test splits} Our retrieval experiments are conducted on a standard ad-hoc IR dataset, the MS MARCO passage collection \cite{msmarco-data}. The relevance of the passages in the MS MARCO collection are more of personalized in nature. A common practice is to use the TREC DL topic sets, which contains depth-pooled relevance assessments on the passages of the MS MARCO collection.
For TREC DL, we conduct experiments on a total of $97$ queries from the years $2019$ and $2020$ \cite{trecdl2020,trecdl2019}. 
Table \ref{tab:dataset} provides an overview of the dataset.

%
%
We use a random sample of $5\%$ of queries (constituting a total of about $40\text{K}$ queries) to train the supervised models in our experiments, whereas evaluation is conducted on the TREC DL (both '19 and '20) query sets.
We use a small sample from the training set since the training process requires executing a feedback model (e.g., RLM) for all queries. Therefore, the model needs to learn a task-specific encoding for each query-document pair, both for the original and the expanded queries.

To investigate the generalisation of our selective feedback model, we employ RLM as the feedback approach to train the decision function (Figure \ref{fig:bert}). During inference, we employ three different PRF approaches, namely RLM, ColBERT-PRF \cite{colbert-prf} and GRF \cite{generative-rlm} to test the effectiveness of selective feedback.





\para{Parameter settings}
A common parameter for all the methods is the number of top-retrieved documents $k$ used for the feedback process and also for training the supervised PRF decision models. 
For each method we tune the $k \in [5,40]$ via grid search on the training folds, and use the optimal value on the test fold. We use the same approach to tune the parameter $\alpha$ in Equation \ref{eq:rankfusion}, which controls the importance of the feedback process for the rank-based fusion methods.
For the \rrf-based methods, we conduct a grid search for $\alpha$ in the set $\{0, 0.1,\ldots, 1\}$. The number of terms used for relevance feedback was tuned for the collection and we use the optimal value across all the methods considered.   

To obtain the initial retrieval list, we use both sparse and dense models. As a sparse model, we employ BM25 \cite{Okapi} to retrieve the top-1000 results from MS MARCO collection and a supervised neural model, namely, MonoT5 \cite{monobert} which operates by reranking the top-1000 of BM25. 
%
MonoT5 model was trained on the MS MARCO training queries.
%

%

\subsection{Results and Discussion} \label{sec:result}

\begin{table*}[t]
\centering
\caption{
\small
Comparison of different SRF approaches on the TREC DL (2019 and 2020) topic sets with BM25 and MonoT5 set as the initial retrieval models.
MAP values are computed for top-1000 documents.
Paired $t$-test ($p<0.05$) shows a significant improvement of \darf~over the best performing baselines (comparing bold-faced results with the underlined ones).
}
\label{tab:trecdl}

\begin{adjustbox}{width=1\textwidth}
\small
\begin{tabular}{@{}llccc|ccc|ccc@{}}

\toprule
& 
& \multicolumn{3}{c}{BM25 ($\phi$: RLM)}
& \multicolumn{3}{c}{BM25 ($\phi$: GRF)}
& \multicolumn{3}{c}{BM25 ($\phi$: ColBERT-PRF)}
\\

\cmidrule(r){3-5}
\cmidrule(r){6-8}
\cmidrule(r){9-11}

& Methods 
& Accuracy & MAP & nDCG@10
& Accuracy & MAP & nDCG@10
& Accuracy & MAP & nDCG@10 \\
 
\midrule

\multirow{6}{*}{Baselines}
& No PRF 
& N/A & 0.3766 & 0.5022
& N/A & 0.3766 & 0.5022
& N/A & 0.3766 & 0.5022 \\

& PRF 
& N/A & 0.4321 & 0.5134
& N/A & 0.4883 & 0.6226
& N/A & 0.4514 & 0.6067 \\

& \rrf
& N/A & 0.4381 & 0.5140
& N/A & 0.5094 & 0.6332
& N/A & 0.4968 & 0.6184 \\

& \qppbert
& 0.7835 & 0.4400 & 0.5152
& \underline{0.7844} & \underline{0.5321} & \underline{0.6667}
& 0.7742 & 0.5238 & 0.6400 \\

& \mcf
& 0.7611 & 0.4392 & 0.5135
& 0.7580 & 0.4579 & 0.5900
& 0.7642 & 0.4910 & 0.6038 \\

& \afr 
& \underline{0.7842} & \underline{0.4411} & \underline{0.5154}
& 0.7784 & 0.5107 & 0.6512
& \underline{0.7854} & \underline{0.5254} & \underline{0.6414} \\

\midrule

\multirow{2}{*}{Ours}
& \darfbert
& \multirow{2}{*}{\textbf{0.8081}$^*$} & 0.4705 & 0.5374
& \multirow{2}{*}{\textbf{0.8093}$*$} & 0.5654 & 0.6821
& \multirow{2}{*}{\textbf{0.8165}$^*$} & 0.5631 & 0.6765 \\

& \darfbert-\rrf
& & \textbf{0.4961} & \textbf{0.5486}
& & \textbf{0.5730} & \textbf{0.6839}
& & \textbf{0.5785} & \textbf{0.6873} \\

\midrule

& Oracle
& \cellcolor{ashgrey}1.0000 
& \cellcolor{ashgrey}0.5038 
& \cellcolor{ashgrey}0.5528 
& \cellcolor{ashgrey}1.0000 
& \cellcolor{ashgrey}0.5876 
& \cellcolor{ashgrey}0.6941
& \cellcolor{ashgrey}1.0000 
& \cellcolor{ashgrey}0.5820 
& \cellcolor{ashgrey}0.6936 \\

\toprule
\toprule

& 
& \multicolumn{3}{c}{MonoT5 ($\phi$: RLM)}
& \multicolumn{3}{c}{MonoT5 ($\phi$: GRF)}
& \multicolumn{3}{c}{MonoT5 ($\phi$: ColBERT-PRF)}
\\

\cmidrule(r){3-5}
\cmidrule(r){6-8}
\cmidrule(r){9-11}

& Methods 
& Accuracy & MAP & nDCG@10
& Accuracy & MAP & nDCG@10
& Accuracy & MAP & nDCG@10 \\

\midrule

\multirow{6}{*}{Baselines}
& No PRF 
& N/A & 0.5062 & 0.6451 
& N/A & 0.5062 & 0.6451
& N/A & 0.5062 & 0.6451 \\

& PRF 
& N/A & 0.5081 & 0.6463
& N/A & 0.5200 & 0.6487
& N/A & 0.5297 & 0.6491 \\

& \rrf
& N/A & 0.5112 & 0.6484
& N/A & 0.5241 & 0.6494
& N/A & 0.5324 & 0.6502 \\

& \qppbert
& \underline{0.7963} & \underline{0.5189} & \underline{0.6559}
& 0.7871 & 0.5313 & 0.6604
& 0.7900 & 0.5419 & \underline{0.6673} \\

& \mcf
& 0.7789 & 0.5071 & 0.6453
& 0.7670 & 0.4991 & 0.6403
& 0.7612 & 0.5179 & 0.5986 \\

& \afr 
& 0.7958 & 0.5180 & 0.6543
& \underline{0.7980} & \underline{0.5422} & \underline{0.6628}
& \underline{0.7928} & \underline{0.5500} & 0.6654 \\

\midrule

\multirow{2}{*}{Ours}

& \darfbert
& \multirow{2}{*}{\textbf{0.8152}$^*$} & 0.5306 & 0.6640
& \multirow{2}{*}{\textbf{0.8160}$*$} & 0.5529 & 0.6694
& \multirow{2}{*}{\textbf{0.8067}$*$} & 0.5624 & 0.6733 \\

& \darfbert-\rrf
& & \textbf{0.5317} & \textbf{0.6659} 
& & \textbf{0.5607} & \textbf{0.6719}
& & \textbf{0.5711} & \textbf{0.6746} \\

\midrule

& Oracle
& \cellcolor{ashgrey}1.0000 
& \cellcolor{ashgrey}0.5416 
& \cellcolor{ashgrey}0.6786
& \cellcolor{ashgrey}1.0000 
& \cellcolor{ashgrey}0.5722 
& \cellcolor{ashgrey}0.6803
& \cellcolor{ashgrey}1.0000 
& \cellcolor{ashgrey}0.5801 
& \cellcolor{ashgrey}0.6821 \\

\bottomrule                  
\end{tabular}
\end{adjustbox}
\end{table*}

\para{Main observations}
The key findings of our experiments are reported in Table \ref{tab:trecdl}. We observe that the accuracy of the decisions is quite satisfactory, even for the unsupervised threshold-based approaches. The results also indicate that more accurate PRF decisions usually lead to an increase in retrieval effectiveness.

For \textbf{RQ1}, we find that supervised selective PRF approaches yield improved results over their unsupervised counterparts. Of particular interest is the fact that a data-driven approach (as per our proposal in this paper) outperforms the feature-based approach \afr~\cite{zhai-cikm09}, which answers RQ1 in the affirmative.

In relation to \textbf{RQ2}, we see that a soft combination of the initial and the feedback lists via a confidence-based calibration (\darfbert-\rrf) improves results further.


\begin{table}[t]
\centering
\caption{
\small
Contingency tables of the \darfbert~model with sample queries from TREC DL. Here, $|Q|$ is the count of queries for each of the 4 possible cases of prediction (true/false positives and true/false negatives), and $\overline{\Delta \text{AP}}$ denotes the average $\Delta \text{AP}$ values of each cell, where $\Delta \text{AP}(Q) = \frac{\text{AP}(\phi_M(Q)) - \text{AP}(Q)}{\text{AP}(Q)}$.
\label{tab:contingency}
}

\begin{adjustbox}{width=.9\columnwidth}
\small
\begin{tabular}{lrlrlr}
 & & \multicolumn{4}{c}{
 \begin{tabular}[c]{@{}c@{}}\bfseries
 Actual\end{tabular}} \\
 
\multicolumn{2}{l}{} 
& \multicolumn{1}{c}{$\Delta \text{AP} >0$} 
&
& \multicolumn{1}{c}{$\Delta \text{AP} \leq 0$}
& \\ 

\cline{3-6} 

\multirow{7}{*}{\begin{tabular}[c]{@{}c@{}}\bfseries \begin{turn}{90}Predicted\end{turn}\end{tabular}} 
& \multicolumn{1}{l|}{$\Delta \text{AP} >0$} 
& \multicolumn{1}{l|}{\cellcolor{magicmint}What is active margin?}
& \multicolumn{1}{l|}{\cellcolor{magicmint}}
& \multicolumn{1}{l|}{\cellcolor{melon}Why is Pete Rose banned}
& \multicolumn{1}{l|}{\cellcolor{melon}} \\ 

& \multicolumn{1}{l|}{}
& \multicolumn{1}{l|}{\cellcolor{magicmint}}
& \multicolumn{1}{l|}{\cellcolor{magicmint}$|Q|=59$}
& \multicolumn{1}{l|}{\cellcolor{melon}from hall of fame?}
& \multicolumn{1}{l|}{\cellcolor{melon}$|Q|=8$} \\ 

\cline{3-3}
\cline{5-5}

& \multicolumn{1}{l|}{}
& \multicolumn{1}{l|}{\cellcolor{magicmint}Exon definition Biology}
& \multicolumn{1}{l|}{\cellcolor{magicmint}$\Delta \overline{\text{AP}}=0.1302$}
& \multicolumn{1}{l|}{\cellcolor{melon}What are best foods to}
& \multicolumn{1}{l|}{\cellcolor{melon}$\Delta \overline{\text{AP}}=0.0525$} \\

& \multicolumn{1}{l|}{}
& \multicolumn{1}{l|}{\cellcolor{magicmint}}
& \multicolumn{1}{l|}{\cellcolor{magicmint}}
& \multicolumn{1}{l|}{\cellcolor{melon}lower cholesterol?}
& \multicolumn{1}{l|}{\cellcolor{melon}} \\ 

\cline{3-6} 

& \multicolumn{1}{l|}{$\Delta \text{AP} \leq 0$} 
& \multicolumn{1}{l|}{\cellcolor{melon}Define BMT medical}
& \multicolumn{1}{l|}{\cellcolor{melon}$|Q|=11$}
& \multicolumn{1}{l|}{\cellcolor{magicmint}Do Google docs auto save?}
& \multicolumn{1}{l|}{\cellcolor{magicmint}$|Q|=19$} \\

\cline{3-3}
\cline{5-5}

& \multicolumn{1}{l|}{}
& \multicolumn{1}{l|}{\cellcolor{melon}Who is Robert Gray?}
& \multicolumn{1}{l|}{\cellcolor{melon}$\Delta \overline{\text{AP}}=0.0246$}
& \multicolumn{1}{l|}{\cellcolor{magicmint}How many sons Robert}
& \multicolumn{1}{l|}{\cellcolor{magicmint}$\Delta \overline{\text{AP}}=0.0737$} \\

& \multicolumn{1}{l|}{}
& \multicolumn{1}{l|}{\cellcolor{melon}}
& \multicolumn{1}{l|}{\cellcolor{melon}}
& \multicolumn{1}{l|}{\cellcolor{magicmint}Kraft has?}
& \multicolumn{1}{l|}{\cellcolor{magicmint}} \\

\cline{3-6} 
 
\end{tabular}
\end{adjustbox}

\end{table}

An interesting finding is that the SRF decision function trained on RLM on a set of queries generalises well not only for a different set of queries (the test set), but also across different feedback models. This suggests that the queries which improve with RLM also improve with other feedback models, such as GRF or ColBERT-PRF. This can be seen from the GRF and the ColBERT-PRF group of results for both BM25 and MonoT5. This entails that the SRF based decision function does not need to be trained for specific PRF approaches, which makes it more suitable to use in a practical setup, affirming \textbf{RQ3}. 



We observe that the best results obtained by our method are close to those achieved by an `oracle'. In the ideal oracle scenario, PRF is applied \textit{only if} the AP of a query is actually improved (i.e., the oracle uses the relevance assessments for the test queries). The fact that the results from  \darfbert~are close to the oracle suggests that further attempts to increase the accuracy of PRF decisions may have little impact on retrieval effectiveness, likely due to saturation effects.

\para{Per-query analysis}
Table \ref{tab:contingency} shows examples of queries from the TREC DL dataset. Firstly, we see that the average differences in the AP values before and after feedback are mostly higher for the green cells, which indicates that the penalty incurred due to queries for which the model (\darfbert) predicts incorrectly is not too high. This also conforms to the fact that at close to $80\%$ accuracy, \darfbert~achieves results close to the oracle. Secondly, a manual inspection of the examples reveals that the queries for which the \darfbert~model correctly decides to apply PRF appear to be those with under-specified information needs, i.e., those queries which are likely to be benefited from enrichment, e.g., the query `what is active margin' as seen from Table \ref{tab:contingency}.

\section{Conclusions and Future Work} \label{sec:conclu}
In this paper, we proposed a selective relevance feedback framework that includes a data-driven supervised neural approach to optimize retrieval effectiveness by applying feedback on queries in a selective fashion.
By testing this approach using multiple PRF models over sparse and dense architectures, we observed that it performs favorably compared to alternative strategies, approaching the performance of an oracle system. 

This work opens the door for interesting future studies. Although our method is effective, it requires executing PRF to gauge result quality. 
Exploring techniques to determine the necessity of the PRF step could reduce computational costs for queries where PRF is ultimately ignored. 
Further work could also examine strategies for predicting the parameters of PRF itself, such as the number of relevant documents.

\subsubsection*{\small \textbf{Acknowledgement}} \small The first and the fourth authors were partially supported by Science Foundation Ireland (SFI) grant number SFI/12/RC/2289\_P2. 

\bibliographystyle{splncs04}
\bibliography{refs}

\end{document}